\documentclass[reprint,prd,aps,amsmath,superscriptaddress,twocolumn,amssymb,nofootinbib]{revtex4}

\usepackage{hyperref}
\usepackage{epsfig}
\usepackage{graphicx}

\newcommand{\bea}{\begin{eqnarray}}
\newcommand{\eea}{\end{eqnarray}}
\newcommand{\beq}{\begin{equation}}
\newcommand{\eeq}{\end{equation}}

\begin{document}

\title{Inflation with large supergravity corrections}

\author{Anupam Mazumdar}
\affiliation{Physics Department, Lancaster University, Lancaster LA1 4YB, UK}
\affiliation{Niels Bohr Institute, Copenhagen, Blegdamsvej-17, Denmark}
\author{Seshadri Nadathur}
\affiliation{Rudolf Peierls Centre for Theoretical Physics, University of Oxford, Oxford OX1 3NP, UK}
\author{Philip Stephens}
\affiliation{Physics Department, Lancaster University, Lancaster LA1 4YB, UK}


\begin{abstract}
It is well known that large Hubble-induced supergravity corrections to the inflaton field can ruin the flatness of the potential, thus creating a tension between slow-roll inflation and supergravity. In this paper we show that it is possible to obtain a {\it cosmologically flat} direction, embedded within the minimal supersymmetric standard model, despite very large super-Hubble corrections. As an illustration, we show that a flat direction which is lifted by an $n=6$ operator matches the current cosmic microwave background data for a wide range of the Hubble parameter, $10^{5}~{\rm GeV}\lesssim H_{inf}\lesssim 10^{8.5}$~GeV. Our approach can be applied to any $F$-term inflationary model.
\end{abstract}

\maketitle


\section{Introduction}
\label{section:introduction}
 
Inflation is perhaps the most popular paradigm for creating the observed initial perturbations in the early universe~\cite{Komatsu:2010fb}. A typical inflationary potential requires a flat direction along which slow-roll inflation can take place. However there are a number of effects beyond the standard model which can lift the flatness of the potential at energies below the Planck scale; for a review on inflation, see~\cite{Mazumdar:2010sa}. One such prominent effect is due to gravity, especially within a supersymmetric model of inflation, known as Hubble-induced supergravity (SUGRA) corrections~\cite{Bertolami:1987xb, Copeland, Binetruy, Dine:1995uk,Dine:1995kz}. These corrections are known for spoiling the flatness of the potential and therefore the success of $F$-term models within supersymmetry, for instance hybrid models~\cite{Dvali:1994ms}; for reviews, see~\cite{Mazumdar:2010sa, Masahide}. 

The inflaton potential generically obtains large corrections from a minimal choice of the K\"ahler potential, $K(\phi^{\dagger}\phi)=\phi^{\dagger}\phi$, where $\phi$ is the inflaton field. This can spoil slow-roll and prematurely end inflation. The source of this correction can arise from a large vacuum energy density present in the early universe. It is well known that besides the inflaton energy density there are many sources which contribute to the total energy density~\cite{Mazumdar:2010sa}.

In this paper, we will show that if inflation is driven near a point of inflection in the potential~\cite{Allahverdi:2006iq,AEGJM,Bueno Sanchez:2006xk,Enqvist:2010vd, Hotchkiss:2011am} it is possible to tame these SUGRA corrections within $F$-term inflation, without invoking any symmetry or ad-hoc choice of K\"ahler potential. We will illustrate this by considering a {\it bottom-up} approach, in an effective field theory of a gauge-invariant flat direction of Minimal Supersymmetric Standard Model (MSSM) which are also lifted by non-renormalizable operators. Within the MSSM there are many such $D$-flat directions---e.g.,~\cite{Dine:1995kz,Gherghetta:1995dv}, for a review see~\cite{MSSM-REV}. Although we consider a particular model of inflation within the MSSM, our analysis can be followed for other supergravity models of inflation.


\section{Supergravity corrections}
\label{section:superHubbleinf}

From a low-energy point of view the flat directions of the MSSM are lifted by the $F$-term of the superpotential. Without loss of generality, let us consider one such $D$-flat direction lifted by a non-renormalizable superpotential term\footnote{For example $\Phi$ could be \emph{udd}, or \emph{LLe}, where \emph{u,~d} correspond to the right-handed squarks, and \emph{L,~e} correspond to the left handed slepton and right handed selectron. With R-parity invariant MSSM, both \emph{udd} and \emph{LLe} are lifted by dimension $6$ operators~\cite{MSSM-REV}.}
\beq
W=\lambda \frac{\Phi^{n}}{M_{P}^{n-3}}\,,
\eeq
where $\Phi$ is flat direction superfield, $\lambda\sim {\cal O}(1)$, and $M_{P}=2.4\times 10^{18}$~GeV. For the rest of the paper we will set $\lambda=1$, because rescaling $\lambda$ simply shifts the VEV of the flat direction, which is perfectly acceptable as long as $\phi$ is below $M_{P}$. The $\phi$ field obtains a soft SUSY breaking mass term $m_{\phi}\sim {\cal O}(100~{\rm GeV})$. Together with the non-renormalizable operator, this gives a potential for $\phi$ which is determined by $n$ and $m_{\phi}$.

In addition, there are many possible contributions to the vacuum energy. It is conceivable that at high energies the universe is dominated by a large cosmological constant arising from a string theory landscape~\cite{Douglas:2006es}. Our own patch of the universe could be locked in a false vacuum within an MSSM landscape~\cite{MN-curvaton,Allahverdi:2008bt}, or there could be hidden sector contributions~\cite{Enqvist:2007tf,Lalak:2007rsa}, or there could be a combination of these effects. For simplicity we may attribute such a vacuum energy to a hidden sector. A hybrid model of inflation~\cite{Linde:1993cn} also provides a source of vacuum energy density during inflation. For the purpose of illustration, let us consider a simple example of the hidden sector superpotential, $$W=M^{2}I,$$ where $M$ is some high scale which dictates the initial vacuum energy density, and $I$ is the superfield. One could also consider:
$$W=\phi(I^{2}-M^{2}).$$ 
Our conclusions remain unchanged and do not depend on what sources the vacuum energy.\footnote{The source of large vacuum energy density can also arise within MSSM as shown 
in Ref.~\cite{Allahverdi:2008bt}.}

The total K\"ahler potential can be of the form \cite{Dine:1995kz,Dine:1995uk}:
\begin{equation}
 K=I^{\dagger}I+\phi^{\dagger}\phi +\delta K, 
 \end{equation}
where the non-minimal term $\delta K$ can be any of these functional forms:
\begin{equation}
\delta K=f(\phi^{\dagger}\phi,I^{\dagger}I)\,,~f(I^{\dagger}\phi\phi)\,,~f(I^\dagger I^\dagger\phi\phi)\,,~f(I\phi^{\dagger}\phi)\nonumber
 \end{equation}
(see also~\cite{Kasuya:2006wf}), though one could also concoct more complicated scenarios. We will always treat the fields $I,~\phi\ll M_{P}$, and we always assume $V(I)$ to dominate over $V(\phi)$. The higher order corrections to the K\"ahler potential are extremely hard to compute. It has been done within a string theory setup~\cite{Berg} but only in very special circumstances, and not for MSSM fields. Therefore, we account for the  uncertainty arising from such corrections by introducing a simple phenomenological coefficient, as we shall discuss below. 

From the low-energy perspective, at the lowest order the effective potential for a $\phi$ field will have only {\it one} intermediate scale, which will be determined by  the Hubble parameter. Since  $H\gg m_{\phi}\sim {\cal O}(100\;\mathrm{GeV})$, the potential is (for the derivation, see~\cite{Dine:1995kz,Dine:1995uk,MSSM-REV})
\begin{eqnarray}
\label{superHubblepot}
V(\phi) = V_c+\frac{c_H H^{2}}{2}{\left\vert \phi \right\vert}^2 - \frac{a_H H}{nM_P^{n-3}}\phi^n\,
+ \frac{\left\vert \phi \right\vert^{2(n-1)}}{M_P^{2(n-3)}}\,,
\end{eqnarray}
where $V_c\approx3H^{2}M_{P}^{2}$, and the second term is the celebrated Hubble-induced mass correction to the inflaton potential~\cite{Bertolami:1987xb,Dine:1995uk,Dine:1995kz}. The coefficient $c_{H}$ depends on the exact nature of the K\"ahler potential and it can also absorb the higher order corrections~\cite{Dine:1995uk}.  This term ruins the flatness, as $m_{\phi}^{2}\ll c_{H}H^{2}$. For large $c_{H}\sim{\cal O}(1)$, the potential is $V\approx 3H^{2}M_{P}^{2}+H^{2}\phi^{2}+...$, which leads to the slow-roll parameter $\eta=M_{P}^{2}V''/V=c_{H}\sim {\cal O}(1)$, and therefore enables the field to roll fast, without allowing sufficient time for the universe to inflate. The third term is the Hubble-induced $A$-term,  where the coefficient $a_{H}$ is dimensionless and of order $\sim c_H$. 

The form of Eq.~(\ref{superHubblepot}) differs from the original MSSM inflation models discussed in Refs.~\cite{Allahverdi:2006iq,AEGJM,AKM,Bueno Sanchez:2006xk}.  In these models it was assumed that $V_{c}=0$, and the soft-supersymmetry breaking mass of the MSSM inflaton was bigger than the Hubble expansion rate during inflation, i.e. $m_{\phi}\sim 100-1000~{\rm GeV}\gg H_{inf}$. Therefore the Hubble-induced SUGRA corrections are small in these models, as was shown explicitly in Ref.~\cite{AEGJM}. However, they suffer from a different problem of fine-tuning between the soft SUSY-breaking terms \cite{AEGJM,Enqvist:2007tf,Lalak:2007rsa}. In the current paper, we have an additional source of vacuum energy density as discussed above, so this condition is not satisfied. As shown in \cite{Hotchkiss:2011am}, such a large vacuum energy density can ameliorate the fine-tuning problem faced by the original models of MSSM inflation. However, that paper did not consider the effect of supergravity corrections, as we do here.

\section{Inflection-point inflation}

In this paper we make the observation that in fact there is a range of field values for which such a potential can be sufficiently flat for inflation to occur. In general such a potential admits a point of inflection which was first analyzed in Refs.~\cite{AEGJM, Bueno Sanchez:2006xk}. The condition for this inflection point to be suitable for inflation is $a_H^2\approx8(n-1)c_H$. We characterize the required fine-tuning by the quantity $\beta$ defined as 
\beq
\label{newbeta}
\frac{a_H^2}{8(n-1)c_H} = 1-\frac{(n-2)^2}{4}\beta^2\,.
\eeq
When $\vert\beta\vert$ is small, a point of inflection $\phi_0$ exists such that $V^{\prime\prime}\left(\phi_0\right) =0$, with
\beq
\label{phi0}
\phi_0 = \left(\sqrt{\frac{c_H}{2(n-1)}} H M_P^{n-3}\right)^{\frac{1}{(n-2)}}\,.
\eeq
In order to simplify our analysis of the motion of $\phi$ in the vicinity of the inflection point, we Taylor expand the potential about the point of inflection $\phi_0$
\bea
V\left(\phi\right)=&&V_0 + \alpha\left(\phi-\phi_0\right) + \frac{\gamma}{6}\left(\phi-\phi_0\right)^3 \nonumber \\ 
\label{Taylorexp}
&&+ \frac{\kappa}{24}\left(\phi-\phi_0\right)^4 + \ldots\,,
\eea
where the following relationships hold \cite{Allahverdi:2006iq, Enqvist:2010vd, Hotchkiss:2011am}:
\bea
\label{V0}
V_0 &=& V_c + \frac{(n-2)^2}{2n(n-1)}c_HH^2\phi_0^2,\\
\label{alpha}
\alpha &=& \frac{(n-2)^2}{4}\beta^2c_HH^2\phi_0+\mathcal{O}(\beta^4),\\
\label{gamma}
\gamma &=& 2(n-2)^2\frac{c_HH^2}{\phi_0},\\
\label{kappa}
\kappa &=& 6(n-2)^3\frac{c_HH^2}{\phi_0^2},
\eea
and $V_c=3H^2M_P^2$ as discussed above.\footnote{Although we retain $n$ in these expressions for generality, for the numerical analysis in this paper we take $n=6$, which is the case for the flat directions $udd$,  $LLe$~\cite{Allahverdi:2006iq}, and the flat direction involving the MSSM Higgses $H_{u}H_{d}$~\cite{Chatterjee:2011qr}. The only {\it renormalizable} inflaton candidate is given in~\cite{AKM}.}

 As can be seen, higher order derivatives fall off as powers of $\phi_0^{-1}$ and so the series can safely be truncated at this point. However, for the parameter values considered in this paper, the fourth-order term $\kappa\left(\phi-\phi_0\right)^4/24$, although small, is not always negligible and hence our analysis differs slightly to that followed earlier in Ref.~\cite{Hotchkiss:2011am}.

The slow-roll parameters are defined by $\epsilon \equiv ({M_P^2}/{2}) \left({V^{\prime}}/{V} \right)^2,~
\eta \equiv M_P^2 \left({V^{\prime \prime}}/{V}\right )$,
and from the form of the potential in eq.~\eqref{Taylorexp} we may write these explicitly as:
\bea
\label{eps}
\epsilon(\phi)&=&\frac{M_P^2}{2 V_0 ^2} \left(\alpha+\frac{\gamma}{2}\left(\phi-\phi_0\right)^2 \right)^2\left(1+\Delta_\epsilon\right)^2 \\
\label{eta}
\eta(\phi)&=&-\frac{\gamma M_P^2}{V_0} \left(\phi_0-\phi\right)\left(1+\Delta_\eta\right)\,,
\eea
where $\Delta_\epsilon$ and $\Delta_\eta$ are small quantities defined as
\bea
\label{deltaeps}
\Delta_\epsilon=\frac{\kappa\left(\phi-\phi_0\right)^3}{6\left(\alpha+\frac{\gamma}{2}\left(\phi-\phi_0\right)^2 \right)}\,,
\Delta_\eta = \frac{\kappa\left(\phi-\phi_0\right)}{2\gamma}\,.
\eea
If the perturbations relevant to the CMB spectrum observed today were generated at a field value $\phi=\phi_\mathrm{CMB}$, the amplitude of the power spectrum and the scalar spectral index are given by:
\bea
\label{Pr}
\mathcal{P}_R&=& \frac{1}{24\pi^2M_P^4}\frac{V_0}{\epsilon\left(\phi_\mathrm{CMB}\right) } \\
\label{ns}
n_s &=& 1+2\eta\left(\phi_\mathrm{CMB}\right),
\eea
where we have used the fact that $\epsilon\ll\vert\eta\vert$ when inflation occurs about an inflection point. Now we use these expressions to compare this model to WMAP data.

\section{Comparison to WMAP data}

\begin{figure}[tbp]
\begin{center}
\includegraphics[scale=0.44]{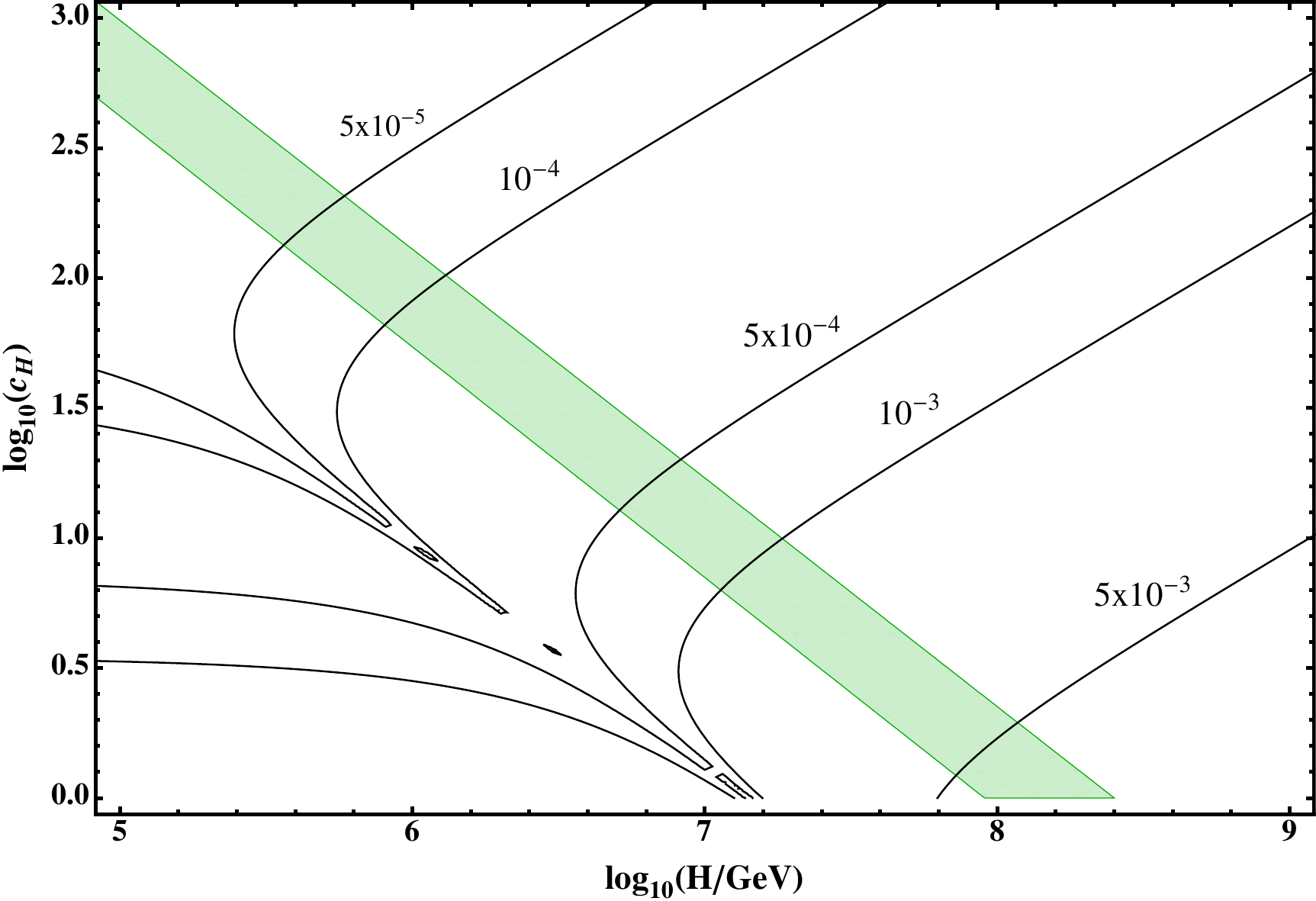}
\caption{Regions of parameter space for the potential in eq.~\eqref{superHubblepot} that satisfy the WMAP 7-year  constraints on the amplitude and spectral index of the power spectrum and also match the e-fold constraint. We minimize $\vert\mathcal{N}-\mathcal{N}_\mathrm{pivot}\vert$ over the range of $\beta$ allowed by the 95\% C.L. constraints on $\mathcal{P}_\mathrm{R}$ and $n_s$. The shaded contour shows the region for which $\left(\vert\mathcal{N}-\mathcal{N}_\mathrm{pivot}\vert\right)_\mathrm{min}\leq1$. Contour lines of $\vert\beta_\mathrm{CMB}\vert$ are shown in black, for the values of $\vert\beta_\mathrm{CMB}\vert$ indicated. We have taken $n=6$. }
\label{figure:c_H-H}
\end{center}
\end{figure}

The WMAP 7-year data suggest a power spectrum with $\mathcal{P}_R=(2.43\pm0.11)\times10^{-9}$ at the pivot scale $k_\mathrm{pivot}=0.002~\mathrm{Mpc}^{-1}$, and a spectral index of $n_s=0.967\pm0.014$ for models with no \textquoteleft running' of the spectral index \cite{Komatsu:2010fb}. Given these constraints eqs.~\eqref{Pr} and \eqref{ns} may be inverted to obtain the values $\epsilon_\mathrm{best}$ and $\eta_\mathrm{best}$ that will produce the required best-fit power spectrum, and the range of $\epsilon$ and $\eta$ values that lie within the $95\%$ confidence limits.

Having obtained the values of $\epsilon_\mathrm{best}$ and $\eta_\mathrm{best}$, one can then in principle invert equations \eqref{eps} and \eqref{eta} to solve for the field value $\phi_\mathrm{CMB}$ and the fine-tuning parameter $\beta$ (or, equivalently, $\alpha$ calculated using equation \eqref{alpha}) for which these values are obtained~\cite{Hotchkiss:2011am}. However in this instance this procedure is complicated by the presence of the terms $\Delta_\epsilon$ and $\Delta_\eta$, so we use an iterative bootstrapping procedure, as follows.

The zeroth-order approximation is $\Delta_{\eta,0}\approx\Delta_{\epsilon,0}\approx0$, so that equation \eqref{eta} can be solved for $\phi_\mathrm{CMB}$,
\beq
\label{phiCMB0}
\phi_{\mathrm{CMB},0}=\phi_0-\frac{V_0\vert\eta_\mathrm{best}\vert}{\gamma M_P^2}\,.
\eeq
Defining $\Delta\phi_{\mathrm{CMB},0}\equiv\left(\phi_{\mathrm{CMB},0}-\phi_0\right)$, we obtain the zeroth-order approximation for $\alpha$,
\beq
\label{alphaCMB0}
\alpha_{\mathrm{CMB},0}=\frac{\sqrt{2\epsilon_\mathrm{best}}V_0}{M_P} - \frac{\gamma}{2}\Delta\phi_{\mathrm{CMB},0}^2\,.
\eeq
Given the $i^\mathrm{th}$order approximation, we can move to $(i+1)^\mathrm{th}$ order using the following equations:
\bea
\label{Deltaeps-i+1}
\Delta_{\epsilon,i+1}&=&\frac{\kappa\Delta\phi_{\mathrm{CMB},i}^3}{6\left(\alpha_{\mathrm{CMB},i}+\frac{\gamma}{2}\Delta\phi_{\mathrm{CMB},i}^2 \right)}\,,\\
\label{Deltaeta-i+1}
\Delta_{\eta,i+1}&=&\frac{\kappa\Delta\phi_{\mathrm{CMB},i}}{2\gamma},\\
\label{phiCMBi+1}
\phi_{\mathrm{CMB},i+1}&=&\phi_0 +\frac{\Delta\phi_{\mathrm{CMB},0}}{\left(1+\Delta_{\eta,i+1}\right)},
\eea
and
\beq
\label{alphaCMBi+1}
\alpha_{\mathrm{CMB},i+1}=\frac{\sqrt{2\epsilon_\mathrm{best}}V_0}{M_P\left(1+\Delta_{\epsilon,i+1}\right)} - \frac{\gamma\Delta\phi_{\mathrm{CMB},i+1}^2}{2}\,.
\eeq
This iterative procedure can be continued to obtain the values of $\phi_\mathrm{CMB}$ and $\alpha_\mathrm{CMB}$ to any desired accuracy. For our purposes, two iterations proved to be sufficient to ensure that further refinement did not produce any noticeable change in the results.
Slow-roll ends at the field value $\phi_{e}$ at which $\vert\eta\vert\sim1$. This can be calculated to the same order of accuracy as the other parameters,
\beq
\label{phieta}
\phi_{e,i}=\phi_0-\frac{V_0}{\gamma M_P^2\left(1+\Delta_{\eta,i}\right)}\,.
\eeq
Inflationary expansion itself will end within a fraction of an e-fold after the field attains the value $\phi_e$. The number of e-folds of inflation produced as $\phi$ rolls from $\phi_\mathrm{CMB}$ to $\phi_e$ is given by $\mathcal{N\left(\phi_\mathrm{CMB}\right)}=\int_{\phi_\mathrm{CMB}}^{\phi_e}\frac{H d\phi}{\dot\phi}$, and can be approximated to the relevant order of accuracy by the numerical integration of
\beq
\mathcal{N}\left(\phi_\mathrm{CMB}\right)=\frac{V_0}{M_P^2}\int_{\phi_{e,i}}^{\phi_{\mathrm{CMB},i}}\frac{\left(1+\Delta_{\epsilon,i}\right)^{-1}d\phi}{\left(\alpha_{\mathrm{CMB},i}+\frac{\gamma}{2}\Delta\phi^2 \right)}\,.
\eeq
An upper bound on the maximum number of e-foldings between the time when the observationally relevant perturbations were generated and the end of inflation can be derived \cite{Liddle:2003as,Hotchkiss:2011am}, under the assumption that the energy scale of inflation is roughly constant during inflation (which is valid as $\epsilon\ll |\eta|\ll 1$):
\beq
\label{Nmax}
\mathcal{N}\leq\mathcal{N}_{\mathrm{pivot}}\equiv64.7+\ln\left(\frac{V_0^{1/4}}{M_P}\right) .
\eeq
At this point, one must ensure that the vacuum energy density which generated the large cosmological constant in the first place vanishes by the end of slow-roll inflation. In the string landscape~\cite{Douglas:2006es}, or in the case of MSSM~\cite{Allahverdi:2008bt}, this can happen through bubble nucleation, provided the rate of nucleation is such that $\Gamma_{nucl}\gg H$. In the latter case all the bubbles will belong to the MSSM vacuum---similar to the first order phase transition in the electroweak symmetry breaking scenario. In the former case, one has to make sure that the cosmological constant disappears in the MSSM vacuum right at the end of inflation~\cite{Allahverdi:2007wh}. The bubble collisions would generate inhomogeneities similar to the electroweak first order phase transition--whose imprints could be found in remnants of high frequency gravitational waves~\cite{Maggiore}. One can as well envisage that there could be a smooth transition of the vacuum energy triggered by the other fields, or possibly by the inflaton itself, similar to the case of hybrid inflation~\cite{Linde:1993cn,Dvali:1994ms}, and as discussed in~\cite{Enqvist:2010vd,Hotchkiss:2011am,Burgess:2005sb}. In either scenario the predictions for the initial seed perturbations for the large scale structures would not be affected.

The time scale for the transfer of energy from $\phi$ to the radiation and the MSSM relativistic species can be computed exactly as in~\cite{Allahverdi:2011aj}. This happens within one Hubble time and thus the inequality in Eq.~\eqref{Nmax} is saturated. This provides a constraint on the model: the values obtained for $\phi_\mathrm{CMB}$ and $\alpha_\mathrm{CMB}$ (or $\beta_\mathrm{CMB}$) from the WMAP7 power spectrum constraints, will only allow $\mathcal{N}\left(\phi_\mathrm{CMB}\right)\approx\mathcal{N}_{\mathrm{pivot}}$ for certain combinations of the free parameters $c_H$ and $H$. 

In Figure~\ref{figure:c_H-H}, we plot the contour of allowed values of $c_H$ and $H$ for which inflation driven by the Hubble-induced corrections can produce a power spectrum of density perturbations consistent with the WMAP7 results, whilst simultaneously satisfying the e-fold constraint $\mathcal{N}\left(\phi_\mathrm{CMB}\right)\approx\mathcal{N}_{\mathrm{pivot}}$ to within an uncertainty of $\pm1$ e-fold. Also plotted are contour lines of the parameter $\vert\beta_\mathrm{CMB}\vert$, which provide an indication of the level of fine-tuning required in equation~\eqref{newbeta}. It can be seen that values of the Hubble parameter $H$ are allowed in the range $10^{5}~{\rm GeV}\lesssim H\lesssim 10^{8.5}$~GeV, where the limits are set simply by the range of values of $c_H$ that we chose to consider. As ${\cal O}(1)\leq c_{H}\leq {\cal O}(10^{3})$, it is seen that the coefficient $c_H$ can take on large values $c_H\gg\mathcal{O}(1)$ without spoiling inflation. This is the main highlight of our paper. Further note that for $c_{H}\sim {\cal O}(1)$, the required fine tuning between $c_{H}$ and $a_{H}$ is less severe, $\beta\sim 10^{-2}$. Interestingly this tuning arises {\it only} during inflation, but will go away at low energies where the predictions for squarks and sleptons will be made at LHC.

\section{Discussion and Conclusion}

A curious reader might wonder what happens if the vacuum energy were made to vanish --- as can be arranged in SUGRA. A very similar plot is obtained when the cosmological constant $V_{c}=0$ in eq.~(\ref{superHubblepot}), while keeping the Hubble-induced mass and the A-term. The parameter space for this case is quite similar to that originally considered in Refs.~\cite{AEGJM,Bueno Sanchez:2006xk}, where the SUGRA corrections are negligible by virtue of $m_{\phi}\gg H$ during inflation. Our numerical findings suggest that we can satisfy the WMAP 7-year constraints for: $c_{H}={\cal O}(1- 10^{4})$ and $H\sim 1-10^{3}$~GeV.

To summarize, we have provided a simple alternative solution to the SUGRA-$\eta$ problem which plagues $F$-term models of inflation. We show that in this case there is no need to make K\"ahler corrections small in order to make the potential flat enough for inflation. Instead, from the low energy point of view it is always possible to find a point of inflection with a VEV below the Planck scale, and a region of field values where the potential is sufficiently flat for successful slow-roll inflation, even in the presence of large K\"ahler corrections. As an offshoot, we also find that the SUGRA corrections can greatly ameliorate the original fine-tuning problem of MSSM inflation~\cite{Allahverdi:2006iq,AEGJM}.

\section{Acknowledgements}

We are thankful to Q. Shafi, M. Yamaguchi, S. Jabbari, and D. Lyth for discussions.

\end{document}